\documentclass[nofootinbib,superscriptaddress,twocolumn,showpacs,preprintnumbers,amsmath,amssymb,prl]{revtex4-1}

\usepackage{graphicx}
\usepackage{grffile}
\usepackage{slashed}
\usepackage{bbm}
\usepackage{footmisc}
\usepackage{multirow}

\usepackage{hyperref}
\usepackage{amssymb}
\usepackage{amsmath}
\usepackage{color}
\usepackage{xspace}
\usepackage{caption}
\usepackage{subcaption}
\usepackage{natbib}

\usepackage{datetime}
\usepackage{color,fancybox}
\newcommand{\GeV}{\ensuremath{\,\mathrm{GeV}}\xspace}
\newcommand{\TeV}{\ensuremath{\,\mathrm{TeV}}\xspace}

\newcommand{\fb}{\ensuremath{\,\mathrm{fb}}\xspace}

\newcommand{\order}[1]{\mathcal{O}\!\left(#1\right)}
\newcommand{\rarrow}{\rightarrow}
\newcommand{\WpZlep}{\ensuremath{e^+ \nu_e \mu^+ \mu^-}}

\newcommand{\bib}[1]{Ref.~\cite{#1}}
\newcommand{\fig}[1]{Fig.~\ref{#1}}

\begin{document}
\title{WZ production in association with two jets at NLO in QCD}

\preprint{FTUV-13-0507\;\; IFIC/13-22\;\; KA-TP-09-2013\;\;LPN13-029\;\;SFB/CPP-13-28}
%\preprint{DCPT/12/140}

\author{Francisco~Campanario}
\email{francisco.campanario@ific.uv.es}
\affiliation{Theory Division, IFIC, University of Valencia-CSIC, E-46980
  Paterna, Valencia, Spain.}
\author{Matthias~Kerner}
\email{matthias.kerner@kit.edu}
\author{Le~Duc~Ninh}
\email{duc.le@kit.edu}
\author{Dieter~Zeppenfeld}
\email{dieter.zeppenfeld@kit.edu}
\affiliation{Institute for Theoretical Physics, KIT, 76128 Karlsruhe, Germany.}

\begin{abstract}
We report on the calculation of $W^\pm Zjj$ production with leptonic 
decays at hadron-hadron colliders at next-to-leading order in QCD. These 
processes are important both to test the quartic gauge couplings of the Standard Model 
and because they constitute relevant backgrounds to 
beyond standard model physics searches. 
Our results show that the next-to-leading order corrections reduce
significantly the scale uncertainties and have a non-trivial phase space dependence. 
\end{abstract}

\pacs{12.38.Bx, 13.85.-t, 14.70.Fm, 14.70.Hp}

\maketitle

%------------------------------------------------------------------------------
The study of di-boson production in association with two jets at the LHC 
is important both to test the quartic gauge couplings of the Standard Model (SM) 
and because they constitute relevant backgrounds to 
beyond standard model physics searches. At leading order (LO), there are two 
distinct production mechanisms. The purely electroweak (EW) contributions of 
the order $\order{\alpha^6}$ include, in particular, the four-vector-boson scatterings 
of the type $VV\to VV$ where the initial gauge bosons are radiated from the 
incoming (anti-)quarks. This ``vector-boson-fusion'' mechanism 
has been considered at next-to-leading order (NLO) in 
QCD for $W^+W^-$~\cite{Jager:2006zc}, 
$ZZ$~\cite{Jager:2006cp}, $W^\pm Z$~\cite{Bozzi:2007ur} 
and the equal-charge $W^+W^+$/$W^-W^-$~\cite{Jager:2009xx,
  Denner:2012dz} production processes. 
In addition, 
there are QCD contributions of the order $\order{\alpha_s^2\alpha^4}$. 
The NLO QCD corrections to these contributions are much more difficult because QCD 
radiation occurs already at LO, leading to complicated topologies with non-trivial 
color structures at NLO. The calculations for $W^+W^-jj$ production have been presented in 
Refs.~\cite{Melia:2011dw, Greiner:2012im} and for the $W^+W^+jj$ case in \bib{Melia:2010bm}.     
In this letter, we present the first theoretical prediction for $W^\pm Zjj$ production 
at order $\order{\alpha_s^3\alpha^4}$. 
The interference effects between the EW and QCD amplitudes are not
considered in this letter and can be neglected 
for a-few-percent precision measurements at the LHC. 
This is justified because those effects are color and kinematically suppressed, 
since the EW and QCD contributions peak in different 
phase-space regions.    

The leptonic decays of the EW gauge bosons are consistently included, with all off-shell effects 
and spin correlations taken into account. The charged leptons in the final state can stem from 
a $Z$ boson or a virtual photon. All possibilities are included to form an EW and QCD gauge invariant set. 
Therein, the dominant contribution comes from the diagrams where 
both $W^\pm$ and $Z$ can be simultaneously on-shell. 
In the following, we consider the specific leptonic final state $e^+\nu_e\mu^+\mu^-$ 
and $e^-\bar{\nu}_e\mu^-\mu^+$. The total results for all possible decay
channels ({\it i.e.} $e^+\nu_e\mu^+\mu^-$, $\mu^+\nu_\mu e^+e^-$, 
$e^+\nu_e e^+e^-$, $\mu^+\nu_\mu \mu^+\mu^-$ in 
the $W^+Zjj$ case and accordingly for the $W^-Zjj$ production) 
can, apart from negligible identical lepton interference effects, 
be obtained by multiplying our predictions by a factor four. 
For simplicity, we choose to describe the resonating $W^\pm$ and $Z$ propagators 
with a fixed width and keep the weak-mixing angle real.

\begin{figure}[h]
  \centering
\includegraphics[width=0.65\columnwidth]{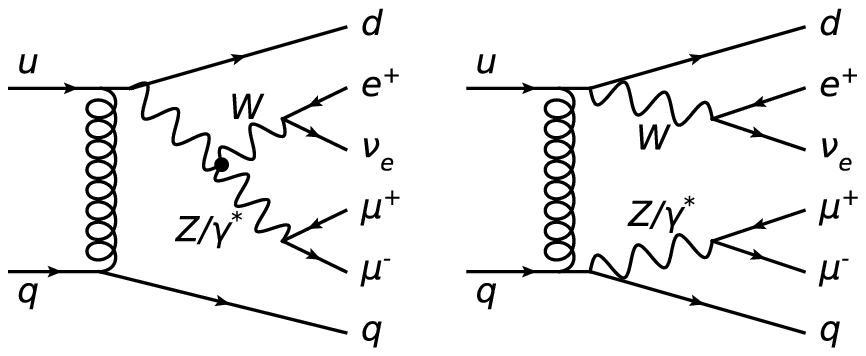}
\includegraphics[width=0.65\columnwidth]{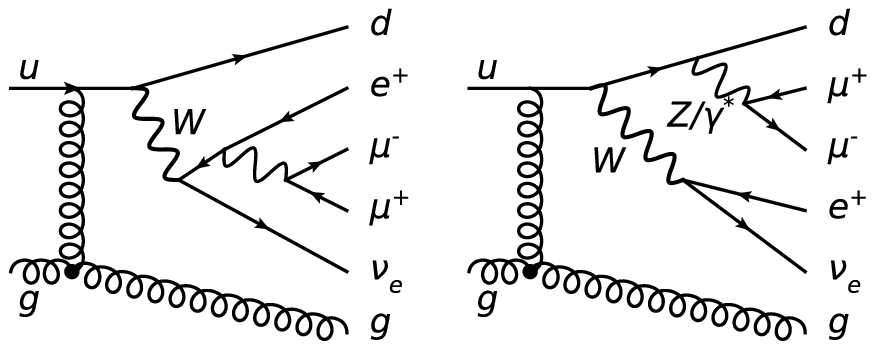}
\caption{Representative tree-level Feynman diagrams.}
\label{fig:feynTree}
\end{figure}
\begin{figure}[th]
  \centering
  \begin{subfigure}[b]{0.27\columnwidth}
    \centering
    \includegraphics[scale=0.65]{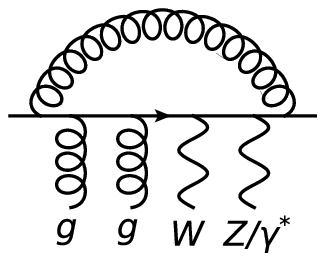}
    \caption{}
    \label{fig:HexLine}
  \end{subfigure} 
  \begin{subfigure}[b]{0.27\columnwidth}
    \centering
    \includegraphics[scale=0.65]{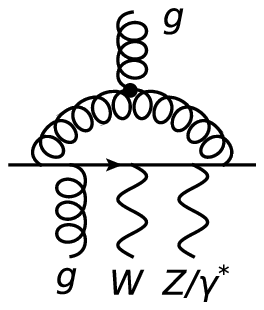}
   \caption{}
  \end{subfigure} 
  \begin{subfigure}[b]{0.27\columnwidth}
    \centering
    \includegraphics[scale=0.65]{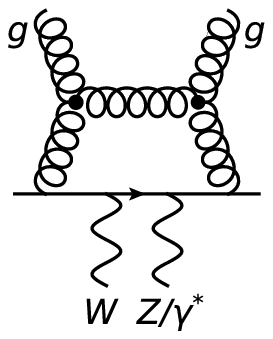}
    \caption{}
  \end{subfigure} 
  \begin{subfigure}[b]{0.27\columnwidth}
    \centering
    \includegraphics[scale=0.65]{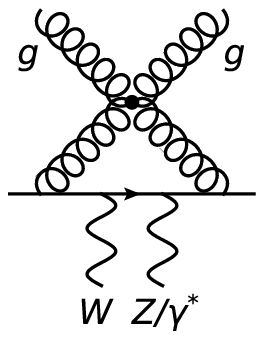}
   \caption{}
  \end{subfigure} 
  \begin{subfigure}[b]{0.27\columnwidth}
    \centering
    \includegraphics[scale=0.65]{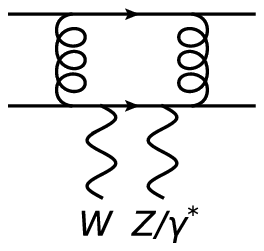}
    \caption{}
  \end{subfigure} 
  \begin{subfigure}[b]{0.27\columnwidth}
    \centering
    \includegraphics[scale=0.65]{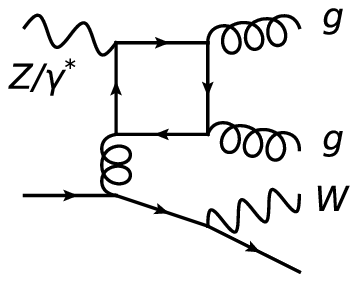}
   \caption{}
  \end{subfigure}
  \caption{Selected Feynman diagrams contributing to the virtual amplitudes.}
  \label{fig:feynVirt}
\end{figure}

The amplitudes are obtained using the Feynman-diagrammatic approach. 
At LO, all contributions to, for example, $pp \rarrow \WpZlep jj$ are classified into 
two groups, $4$-quark and $2$-quark-$2$-gluon amplitudes, as depicted in \fig{fig:feynTree}. 
Each group is then further divided into two QCD-gauge-invariant sub-groups 
(i) with two EW gauge bosons coupling to the quark lines 
and (ii) with the $W^\pm$ radiated from 
a quark line decaying into four leptons. Crossing symmetry is used to obtain all $90$ subprocesses from the minimal set 
of five generic subprocesses, for two generations of quarks. At NLO, there are the virtual and the real 
corrections. \fig{fig:feynVirt} shows some selected contributions to the virtual amplitude, which involves 
in particular the hexagon diagrams. The most difficult part 
of the calculation is computing the $2$-quark-$2$-gluon virtual
amplitudes with up to six-point rank-five one-loop 
tensor integrals. There are $84$ six-point diagrams for each of seven independent subprocesses. 
The $4$-quark group is much easier with only $12$ hexagons for the most complicated subprocesses 
with same-generation quarks. 
The calculation of tensor 
integrals is done by using Passarino-Veltman reduction~\cite{Passarino:1978jh} 
for up to $4$-point diagrams and the method of \bib{Denner:2005nn} (see also \bib{Binoth:2005ff}) for higher-point tensor integrals. 
The scalar integrals are calculated as in Refs.~\cite{'tHooft:1978xw, Dittmaier:2003bc, Denner:2010tr}. 
The real emission contribution includes, for two generations of quarks, $146$ subprocesses with seven particles in the final state. 
Both the virtual and real corrections are, apart from the 
UV divergences in the virtual amplitude which are removed by the renormalization of $\alpha_s$, 
separately infrared divergent. These divergences cancel in the sum for 
infrared-safe observables such as the inclusive cross section and jet distributions.  
We use the dimensional regularization method~\cite{'tHooft:1972fi} 
to regularize the UV and the infrared divergences and apply the 
Catani-Seymour dipole subtraction algorithm~\cite{Catani:1996vz} to combine the virtual 
and the real contributions.   

We have constructed two independent implementations of the above described method. 
The results of the two computer codes are in full agreement, typically $8$ to $12$ digits 
with double precision, at the amplitude level for all subprocesses at NLO. The integrated part of the 
dipole subtraction term in~\bib{Catani:1996vz} has been compared at the integration level. 
Moreover, we have also simplified the calculation to compare with {\texttt{MCFM}}~\cite{Campbell:2002tg,Campbell:2003hd} for the 
case of $e^+\nu_e jj$ production at NLO and found agreement at the per
mille level. 
The first implementation is done in the {\texttt{VBFNLO}} framework~\cite{Arnold:2008rz,*Arnold:2012xn}, which will be described below. 
The second implementation uses {\texttt{FeynArts-3.4}}~\cite{Hahn:2000kx} 
and {\texttt{FormCalc-6.2}}~\cite{Hahn:1998yk} to obtain the virtual amplitudes. 
The scalar and tensor one-loop integrals are evaluated with the in-house 
library {\texttt{LoopInts}}. The tree-level amplitudes for both LO and NLO real emission contributions 
are calculated in an optimized way by using {\texttt{HELAS}}~\cite{Murayama:1992gi,Alwall:2007st} routines.    

In the following, we sketch the main implementation which has been added to the 
{\texttt{VBFNLO}} program and will be made public. 
We use the spinor-helicity formalism of \bib{Hagiwara:1988pp}. 
The virtual amplitudes are constructed using generic building blocks, 
which include sets of loop corrections to Born topologies with a fixed number and a fixed order of external particles. 
For example, by starting from the diagram in \fig{fig:HexLine}, we can form a building block 
by attaching the virtual gluon in all possible ways to the quark line, thereby including also 
pentagon, box, triangle and self energy corrections. 
Within the building blocks, an appropriate color factor is assigned to each Feynman diagram.
Because the building blocks assume the polarization vector of the external 
gauge bosons as an effective current and do not use special properties like 
transversality or being on-shell, they can be also used to check various identities 
relating $N$-point integrals to lower point integrals by replacing a polarization vector 
with the corresponding momentum. Those identities are called gauge tests and 
are checked for every phase space point with a small additional computing cost 
by using a cache system. This is important because the phase space integration 
of the virtual contribution shows numerical instabilities 
in the calculation of one-loop tensor integrals. 
If a bad phase-space point is identified, {\it i.e.} the gauge tests 
are true by less than $2$ digits with double precision, the program 
then calculates the associated building blocks again with quadruple 
precision. The gauge tests are applied again and the point is discarded 
if they fail. For a typical calculation with the inclusive cuts specified below the 
number of discarded points is statistically negligible. 
This strategy was also successfully applied for $W\gamma \gamma$ + jet at
NLO  QCD in \bib{Campanario:2011ud}. 
Further details about the building blocks, tensor reduction master equations and the issue of numerical instabilities 
can be found in \bib{Campanario:2011cs} and also in a forthcoming publication. 
Since the leptonic decays of the EW gauge bosons are common for all subprocesses, 
the {\texttt{VBFNLO}} approach is to calculate these decays once for 
each phase-space point and store them. 
Due to the large number of subprocesses, 
we extend this procedure and also precalculate parts of Feynman diagrams, that are common to the subprocesses of the real emission. 
In addition, a caching system to reuse Born amplitudes for 
different dipole terms~\cite{Catani:1996vz} has been implemented. 
With this method, we obtain the NLO inclusive cross section with statistical error of $1\%$ in $2.5$ hours 
on an Intel $i5$-$3470$ computer with one core and using the compiler Intel-ifort version $12.1.0$.

We use $M_Z=91.1876 \GeV$, $M_W=80.385 \GeV$ and
$G_F=1.16637\times 10^{-5}\GeV^{-2}$ as EW input parameters and 
derive the weak-mixing angle from the SM tree-level relations. 
All fermions but the top quark are approximated as massless.
The widths are calculated as $\Gamma_{W}=2.09532 \GeV$, $\Gamma_{Z}=2.50606 \GeV$. 
The MSTW2008 parton distribution functions (PDF)~\cite{Martin:2009iq} 
with $\alpha_s^\text{LO}(M_Z)= 0.13355$ and $\alpha_s^\text{NLO}(M_Z)= 0.11490$ are used. 
The numerical results presented in this letter are calculated in the four-flavor scheme 
({\it i.e.} the third-generation quarks are excluded in the calculation of 
the $\alpha_s$ running, the PDF evolution and the amplitudes of the 
hard processes) for the LHC at $14 \TeV$ center-of-mass energy. 
This choice is justified because the external $b$-quark contribution 
can, in principle, be excluded by using $b$ tagging. The results for the five-flavor scheme 
will be presented elsewhere. 
Effects from generation mixing are neglected. 
To have a large phase space for QCD radiation, we choose inclusive cuts. 
The charged leptons are required to be hard and central: $p_{T,\ell}\ge 20 \GeV$ 
and $|y_{\ell}|\le 2.5$. The missing transverse energy must
satisfy the cut $E_{T,\text{miss}} \ge 30 \GeV$. For any pair of 
opposite-charge leptons, we impose $m_{\ell^+\ell^-} \ge 15 \GeV$, 
which avoids collinear singularities coming from off-shell photons $\gamma^*\to \ell^+ \ell^-$. 
All final state partons are clustered 
to jets by using the anti-$k_t$ algorithm~\cite{Cacciari:2008gp} with the radius $R=0.4$. 
There must be at least two hard jets with $p_{T,\text{jet}}\ge 20\GeV$ and $|y_{\text{jet}}|\le 4.5$. 
In addition, we impose a requirement on the lepton-lepton and lepton-jet
separation in the azimuthal angle-rapidity plane $\Delta R_{\ell(\ell,j)} \ge 0.4$, where 
only jets passing the above cuts are involved. 
As the central value for the factorization and renormalization scales, we choose
$\mu_{F}=\mu_{R}=\mu_{0}=
\left(\sum_\text{jet}  p_{T,\text{jet}} + 
\sqrt{p_{T,W}^2+m_W^2} +
\sqrt{p_{T,Z}^2+m_Z^2}\right)/2
$,
where $p_{T,V}$ and $m_V$ with $V$ being $W$ or $Z$ are understood as the reconstructed
transverse momenta and invariant masses of the decaying bosons and the sum 
includes only jets passing all cuts.

\begin{figure}[t]
  \centering
  \includegraphics[width=0.83\columnwidth]{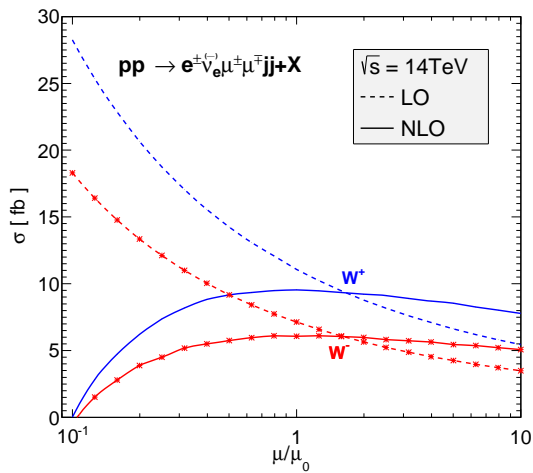}
\caption{Scale dependence of the LO and NLO cross sections at the LHC. 
The curves with and without stars are for $W^-Zjj$ and $W^+Zjj$ productions, 
respectively.}
\label{fig:scale}
\end{figure}
Since the results are calculated at a fixed order 
in perturbative QCD, they depend on the arbitrary scales $\mu_F$ and $\mu_R$. 
The validity of the theoretical predictions is established by proving 
that the scale dependence reduces when higher-order terms are included. 
This is shown in \fig{fig:scale} for the cross section calculated at LO and NLO. 
\begin{figure}[t]
  \centering
  \includegraphics[width=0.83\columnwidth]{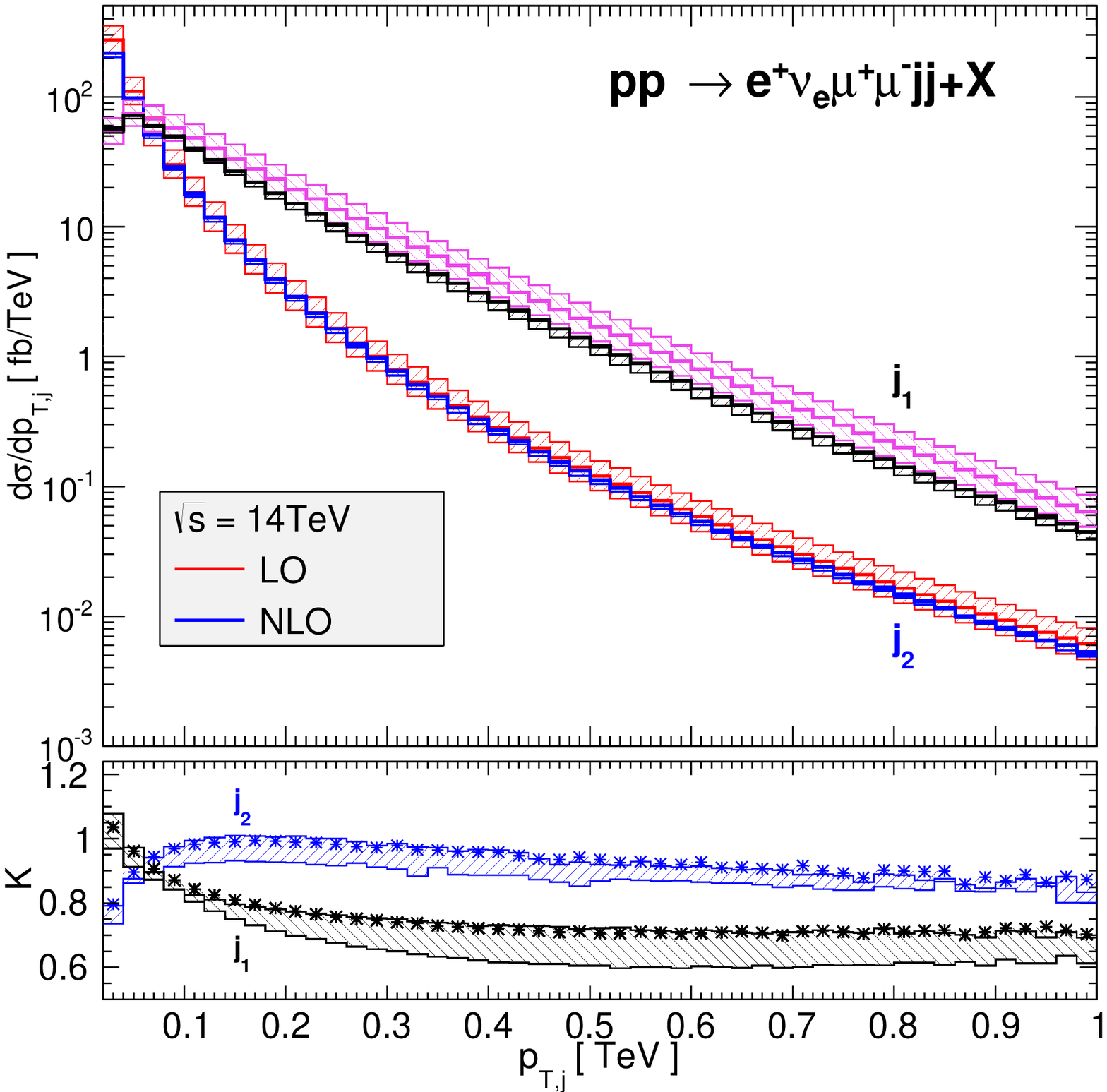}
  \includegraphics[width=0.83\columnwidth]{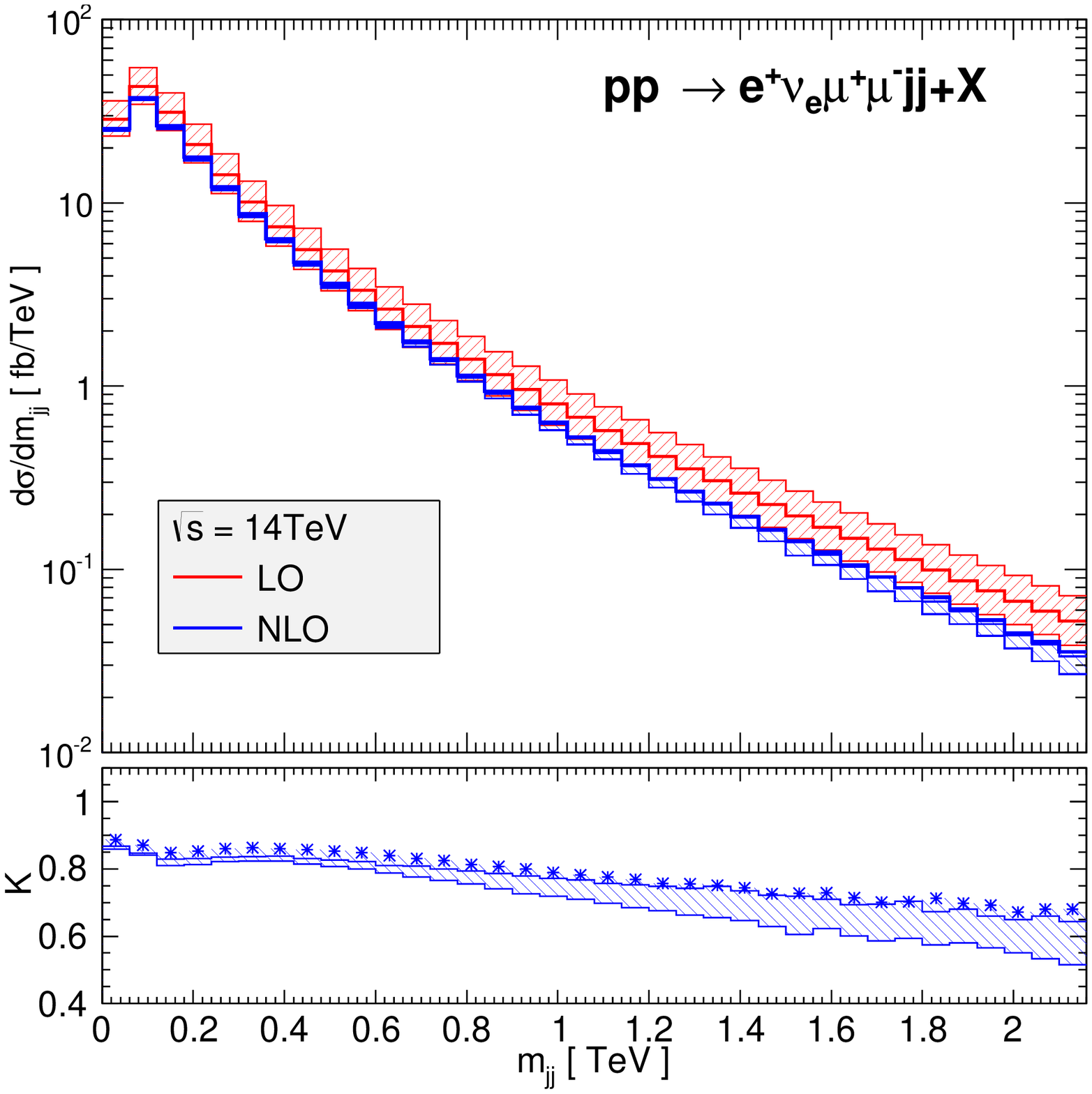}
  \caption{Differential cross sections and K factors for the transverse momenta (top) and 
the invariant mass (bottom) of the two hardest jets. 
The bands describe $\mu_0/2 \le \mu_F=\mu_R\le 2\mu_0$ variations. 
The $K$-factor bands are due to the scale variations of the NLO results, 
with respect to $\sigma_\text{LO}(\mu_0)$. 
The curves with stars in the lower panels are for the central scale, while the two solid lines correspond to $\mu_F = \mu_R=2\mu_0$ and $\mu_0/2$.}
\label{fig:dist_jet}
\end{figure}
For simplicity, the two scales are set equal. As expected, 
we observe a significant reduction in the scale dependence around the central value 
$\mu_0$ when the NLO contribution is included. For both $W^+$ and $W^-$ cases, the uncertainties 
obtained by varying $\mu_{F,R}$ by factors $1/2$ and $2$ around the central value 
are $50\%$ at LO and $5\%$ at NLO. At $\mu = \mu_0$, we get $\sigma_\text{LO}=11.1^{+3.2}_{-2.3}\fb$($7.1^{+2.0}_{-1.5}\fb$) 
and $\sigma_\text{NLO}=9.5^{+0.0}_{-0.4}\fb$($6.1^{+0.0}_{-0.3}\fb$) for the $W^+$($W^-$) case. 
At LO, the dominant contribution is from the $2$-quark-$2$-gluon group, about $86\%$ for both cases. 
By varying the two scales separately, we observe a small dependence on $\mu_F$, while 
the $\mu_R$ dependence is similar to the behavior shown in \fig{fig:scale}. 

\begin{figure}[h]
  \centering
  \includegraphics[width=0.83\columnwidth]{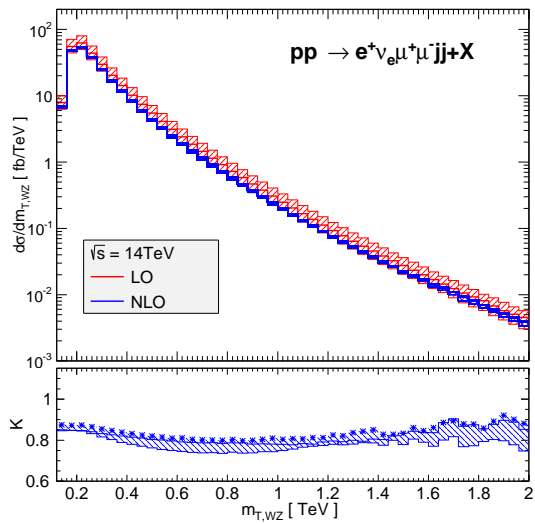}
  \caption{Similar to \fig{fig:dist_jet} but for 
the transverse mass of the two gauge bosons $m_{T,WZ}$. 
}
\label{fig:dist_wz}
\end{figure}
We show the distributions of the 
transverse momenta and 
the invariant mass of the two hardest jets in \fig{fig:dist_jet}, and 
the transverse mass of the two gauge bosons $m_{T,WZ}$ in \fig{fig:dist_wz}. 
We define, as in \bib{Englert:2008tn}, $m_{T,WZ}^2 = (\sqrt{m^2(\ell\ell\ell)+p^2_T(\ell\ell\ell)} + 
\vert p_{T,\text{miss}} \vert )^2 - 
(\vec{p}_T(\ell\ell\ell) + \vec{p}_{T,\text{miss}})^2$, 
with $m(\ell\ell\ell)$ and $p_T(\ell\ell\ell)$ denoting the invariant 
mass and transverse momentum of the charged-lepton system, respectively. 
The $K$ factors, defined as the ratio of the NLO to the LO results, are 
shown in the lower panels. The distributions at NLO are much less sensitive
to the variation of the scales than at LO. The $K$ factors vary from $0.6$ to $1$ 
in a large energy range. This fact together with \fig{fig:scale} indicate that 
we should choose a larger central scale, about $2\mu_0$, to bring the 
LO results closer to the NLO ones for the inclusive cuts. 
We have also studied a fixed scale choice such as $\mu_0^\text{fix} = 400\GeV$ and found that 
the NLO inclusive cross section as a function of the scales is stable around $\mu_0^\text{fix}$ 
and is close to the LO one as well as the dynamic scale prediction. However, the 
transverse momentum and the invariant mass distributions become unstable at large $p_{T}$, with 
very small $K$ factors. This is because the bulk of the inclusive cross section comes from the low energy regime 
as shown in \fig{fig:dist_jet} and \fig{fig:dist_wz}, but a fixed energy scale is not appropriate for all energy regimes.   
The steep increase of the $K$ factor for the transverse momentum distribution of the second hardest jet 
near $20\GeV$ is probably a threshold effect: the phase space for three-visible-jet events is opened up 
as $p_{T,j_2}$ grows well above the cut of $20\GeV$. 

In this letter, we have reported on the first calculation of $W^{\pm}Zjj + X$ production at order $\order{\alpha_s^3 \alpha^4}$ and found K factors close to one. While further phenomenological results will be presented in a future paper, we also plan to make the code publicly available as part of the {\texttt{VBFNLO}} program~\cite{Arnold:2008rz,*Arnold:2012xn}.

We acknowledge the support from the Deutsche Forschungsgemeinschaft
via the Sonderforschungsbereich/Transregio SFB/TR-9 Computational Particle Physics.
FC is funded by a Marie Curie fellowship (PIEF-GA-2011-298960) and partially by MINECO (FPA2011-23596) and by LHCPhenonet (PITN-GA-2010-264564).  
MK thanks the ``Strukturiertes Promotionskolleg in KCETA'' for financial support.

%%%%%%%%%%%%%%%%%%%%%%%%%%%%%%%%%%%%%%%%%%%%%%%%%%%%%%%%%%%%%%
%\bibliographystyle{h-physrev}
%\bibliography{QCDWZjj_short}

\end{document}